\documentclass[final,5p,times,twocolumn,authoryear]{elsarticle}

\usepackage[english]{babel}
\usepackage{amsmath,amssymb,amsfonts}
\usepackage{algorithm}
\usepackage{algorithmic}
\usepackage{graphicx}
\usepackage{tikz}
\usetikzlibrary{shapes,arrows,calc,positioning}
\usepackage{textcomp}
\usepackage{dsfont}
\usepackage{xcolor}
\usepackage{units}
\usepackage{booktabs}
\usepackage{comment}
\usepackage{url}
\usepackage[normalem]{ulem}
\usepackage[hidelinks]{hyperref}
\usepackage{caption}
\usepackage{subcaption}
\usepackage{flushend}
\usepackage{lettrine}
\usepackage{lipsum}

\newcounter{thm}
\newtheorem{prob}[thm]{Problem}

\newtheorem{theorem}{Theorem}[section]

\newif\ifmargincomments
\margincommentstrue % Comment  this line out to turn off comments

\newif\ifrev
\revtrue % Comment  this line out to unmark revisions

\ifmargincomments

\else

\fi

\ifrev

\else

\fi

\newcommand{\flexbrac}[1]{\if\relax\detokenize{#1}\relax \else (#1) \fi}
\newcommand{\flexcomma}[1]{\if\relax\detokenize{#1}\relax \else ,#1 \fi}

\newcommand{\abs}[1]{\lvert #1 \rvert}
\newcommand{\cA}{\mathcal{A}}

\newcommand{\cC}{\mathcal{C}}

\newcommand{\cG}{\mathcal{G}}

\newcommand{\cM}{\mathcal{M}}

\newcommand{\cV}{\mathcal{V}}

\newcommand{\sN}{\mathbb{N}}

\journal{CEP}

\begin{document}
	
\begin{frontmatter}
	\title{Ride-pooling Electric Autonomous Mobility-on-Demand: Joint Optimization of Operations and Fleet and Infrastructure Design}
	\author{Fabio~Paparella\fnref{addressTUe}\corref{correspondingAuthor}} 
	\ead{f.paparella@tue.nl}
	\author{Karni~Chauhan\fnref{addressTUe}}
	\author{Luc~Koenders\fnref{addressTUe}}
	\author{Theo~Hofman\fnref{addressTUe}}
	\author{Mauro~Salazar\fnref{addressTUe}}
	\address[addressTUe]{Control Systems Technology section, Eindhoven University of Technology, the Netherlands}
	\cortext[correspondingAuthor]{Corresponding author.}
	
	\begin{abstract}             
This paper presents a modeling and design optimization framework for an Electric Autonomous Mobility-on-Demand system that allows for ride-pooling, i.e., multiple users can be transported at the same time towards a similar direction to decrease vehicle hours traveled by the fleet at the cost of additional waiting time and delays caused by detours.
In particular, we first devise a multi-layer time-invariant network flow model that jointly captures the position and state of charge of the vehicles.
Second, we frame the time-optimal operational problem of the fleet, including charging and ride-pooling decisions as a mixed-integer linear program, whereby we jointly optimize the placement of the charging infrastructure.
Finally, we perform a case-study using Manhattan taxi-data.
Our results indicate that jointly optimizing the charging infrastructure placement allows to decrease overall energy consumption of the fleet and vehicle hours traveled by approximately 1\% compared to an heuristic placement.
Most significantly, ride-pooling can decrease such costs considerably more, and up to 45\%. 
Finally, we investigate the impact of the vehicle choice on the energy consumption of the fleet, comparing a lightweight two-seater with a heavier four-seater, whereby our results show that the former and latter designs are most convenient for low- and high-demand areas, respectively.
	\end{abstract}
\end{frontmatter}

%===============================================================================

\section{Introduction}\label{sec:Intro}
Recent advances in autonomous driving and powertrain electrification are paving the way to the deployment of Electric Autonomous Mobility-on-Demand (E-AMoD) systems as shown in~Fig.~\ref{fig:Overview}, whereby electric self-driving robo-taxis provide on-demand mobility. %This system allows users to use a mobility service \msmargin{without the need to own a car}{e quando sarebbe servita?}. However, it has been shown that replacing \msmargin{private vehicles}{with what?} could lead to an increase in vehicle-hours traveled (VHT), mainly due to empty-mileage trips, as shown by~\cite{DiaoKongEtAl2021}. 
In this context, both operational aspects such as vehicle routing, rebalancing, ride-pooling and charge scheduling, and design aspects such as vehicle design in terms of battery capacity and number of seats, and charging infrastructure placement, have a strong impact on the achievable performance, whilst also being intimately coupled:
\begin{figure}[t!]
		\centering
		\includegraphics[trim={0 50 0 0},clip,width=0.7\columnwidth]{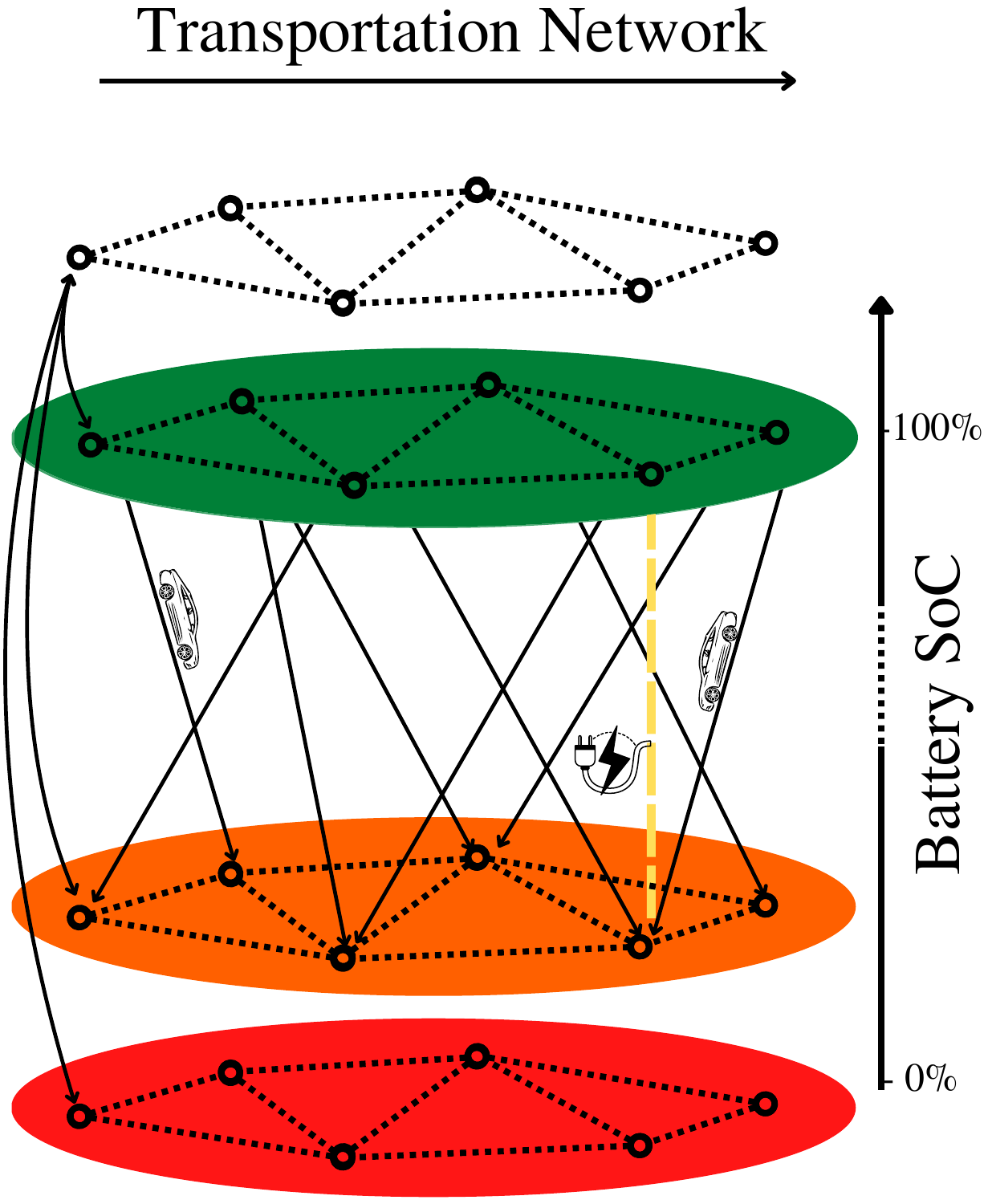} \caption{Multi-layer network schematically representing an E-AMoD system. Each layer corresponds to a battery state of charge (SoC). The top layer represents the original network where each node, called geo-node, represents a geographical location. The curved geo-arcs connect every layer to the same geographical location. As a car travels, it drives through the solid arcs and decrease its SoC. The yellow arcs are placed at charging stations and capture charging activities. }
		\label{fig:Overview}
\end{figure}
%Thus, it is important to not only deploy an efficient systems in terms of  charging infrastructure, vehicle design and operation, %For this reason, it is crucial not only to build incentive schemes to steer the behavior of people towards societal objectives~\citep{PaccagnanChandanEtAl2019,PedrosoHeemelsEtAl2023}, 
%but also to allow and promote ride-pooling, i.e., allowing a vehicle to transport two or more users at the same time towards a similar direction,\msmargin{ leveraging incentives ~\citep{FielbaumKucharskiEtAl2022,PaccagnanChandanEtAl2019,PedrosoHeemelsEtAl2023}.}{beyond the scope?}  These aspects have the potential to directly decrease VHT, and in turn emissions and congestion. %, at the cost of an additional delay and waiting time experienced by users. 
%The ride-pooling problem is notoriously combinatorial~\citep{Santi2014}, and it is difficult to solve, especially in conjunction with other problems like charging infrastructure siting and \msmargin{rebalancing}{it is part of the rp problem}.
For instance, the battery capacity of the vehicles and the charging infrastructure placement influence the fleet operation in terms of routing and charge scheduling, whilst the number of seats constrain the implementable ride-pooling schemes and dictate the energy consumption, which in turn, influence the vehicle hours traveled (VHT) and charging schedules.
Therefore, all the aforementioned design and operational aspects should be jointly optimized when deploying E-AMoD systems.
This paper proposes an optimization framework to optimize the planning of an E-AMoD system that allows for ride-pooling, jointly with the placement of the charging infrastructure, whilst minimizing the fleet size in a computationally efficient and tractable manner.
The proposed framework allows to rapidly explore a large design space, which is a desirable feature when comparing, for example, different vehicle's architectures and fleet compositions.

\textit{Related Literature:} 
This paper pertains to the research streams of AMoD operation, ride-pooling, and charge scheduling and charging infrastructure siting.

A multitude of approaches to model and control AMoD systems are available~\citep{BanerjeeJohariEtAl2015,IglesiasRossiEtAl2017,LevinKockelmanEtAl2017,HoppeEndersEtAl2024,TuranTuckerEtAl2019,AlizadehWaiEtAl2014,LeFlochMeglioEtAl2015}.
In particular, multi-commodity network flow models~\citep{SpieserTreleavenEtAl2014,Pavone2015} are ideal for design optimization purposes as they do not scale with the number of vehicles, and allow for the implementation of multiple objectives and constraints, such as the minimization of VHT by the fleet.

Ride-pooling was studied in recent years. In~\cite{Santi2014}, the authors analyzed the benefits of pooling, while in~\cite{Alonso_Mora_2017} they developed an algorithm that can solve the ride-pooling problem in an optimal manner with high capacity vehicles. However, these microscopic frameworks are computationally very expensive, and cannot be directly leveraged for design optimization purposes.
Against this backdrop, we  recently proposed a linear network flow modeling approach that allows to study ride-pooling AMoD from a mesoscopic planning perspective~\citep{PaparellaPedrosoEtAl2024b}, also accounting for endogenous congestion~\citep{PaparellaPedrosoEtAl2024}, but not including electric vehicles.

Finally, the deployment of electric vehicles introduce the energy consumption and charging dimension, and was studied with directed acyclic graphs in~\cite{BoewingSchifferEtAl2020,PaparellaHofmanEtAl2024}, which can also be leveraged for other electric mobility problems such as freight transportation~\citep{WangZengEtAl2023,BertucciHofmanEtAl2024} and electric air mobility~\citep{VehlhaberSalazar2023b}.
Mescoscopic planning models were studied accounting for the interaction with the power grid in~\citep{RossiIglesiasEtAl2018b,EstandiaSchifferEtAl2021}, whilst a time-varying E-AMoD model was proposed in~\cite{LukeSalazarEtAl2021} that allows to jointly optimize the operations and charging infrastructure for an E-AMoD system, however at the cost of large computational efforts for relatively small and clustered road networks, and without accounting for ride-pooling.
%Having a tractable problem can be extremely important, especially when the goal is to explore different parametrizations, including combinatorial problems like ride-pooling, and perform comparison studies, e.g., in terms of vehicular composition of the fleet.

In conclusion, to the best of the authors' knowledge, there are no mesoscopic models available to optimize the operations of an E-AMoD fleet with ride-pooling, jointly with the charging infrastructure siting in a computationally-tractable manner and with global optimality guarantees.

\textit{Statement of Contributions:} This paper presents a modeling and optimization framework to jointly optimize the ride-pooling and charging operations of an E-AMoD fleet jointly with the charging infrastructure placement in a computationally-effective manner and with global optimality guarantees. 
First, we propose a linear time-invariant network flow model capturing the fleet routing including ride-pooling and charging activities, and combine it with the combinatorial infrastructure design problem. 
%We perform a smart pruning of the road network that allows to generate a synthetic . 
Second, we frame the optimization problem as a mixed-integer linear program that can be efficiently solved with off-the-shelf optimization algorithms. 
Finally, we showcase our framework with two case studies in Manhattan, NYC, USA.
%The first case study shows the impact of the siting and density of the charging infrastructure on the energy consumed by the user-free flow of vehicles, and the second explores the impact of a given vehicle on the fleet sizing and energy consumption.

A preliminary version of this paper was presented at
the 2023 IFAC World Congress~\citep{PaparellaChauhanEtAl2023}.
In this extended version, we include the ride-pooling capabilities of the framework, that allow to study the performance of the system as a function of the maximum waiting time and delay experienced by the users.
In addition, we develop a pruning algorithm that allows to transform a road network in a synthetic road network with iso-energy consumption arcs, which enables a significant decrease in the computational complexity of the problem.
Finally, we study the performance achievable by different electric vehicles.
   
\textit{Organization:} The remainder of this paper is structured as follows: Section~\ref{sec: Methodology} devises the E-AMoD modeling and optimization framework. 
Section~\ref{sec: Results} presents our case studies for Manhattan, NY.
Finally, Section~\ref{sec: Conclusions} draws the conclusions and provides an outlook on future research.

\section{Methodology}\label{sec: Methodology}
In this section, we present a multi-layer time-invariant network flow model to optimize the operation of an E-AMoD system with ride-pooling jointly with the placement of the charging infrastructure, Fig.~\ref{fig:Overview}.
%\msmargin{Given the multi-layer network shown in , a node in the network represents both the position of vehicles on the road network and their state of charge (SoC).}{e...? come mai qua?}
 
\subsection{Multi-layer Network} 
We model the transportation network as a directed graph $\mathcal{G_\mathrm{R}} = (\mathcal{V_\mathrm{R}}, \mathcal{A_\mathrm{R}})$ with a set of nodes $\mathcal{V_\mathrm{R}}$ representing the location of intersections on the road network, and  a set of arcs $\mathcal{A_\mathrm{R}}$ representing the road link between connected nodes. Each road arc $(i,j) \in \mathcal{A_\mathrm{R}}$ is characterized by a distance $d_{ij}$, travel time $t_{ij}$, and energy  $e_{ij}$ required to traverse it.

\begin{figure}[t]
\includegraphics[width=\columnwidth]{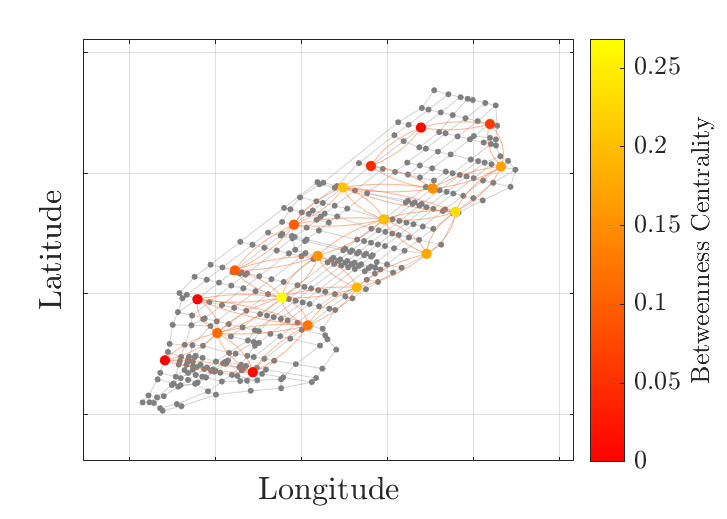}
\caption{Original road network of Manhattan (grey). Pruned synthetic network (colored) with equal (energy) weights.}
\label{fig:FinalGraphs}
\end{figure}
To model the SoC of the fleet, we build a multi-layer network $\mathcal{G} = (\mathcal{V}, \mathcal{A})$, where each layer has the same set of nodes of the original road network $\mathcal{V_\mathrm{R}}$. Fig.~\ref{fig:Overview} shows a schematic representation of the multi-layer network. Each node in each layer has a set of "copy nodes" in the other layers that represent the same geographical location, but indicate a different SoC. The first layer indicates the battery at $100$\%, the last layer at $0$\% SoC, and the number of layers $n$ depends on the discretization of the SoC of the battery. We define the set of nodes of the multi-layer network as the union of the set of nodes in every layer, $\cV  = \cV_1 \cup \cV_2 \cup .. \cV_n$. As vehicles commute each arc, they also move to a lower layer, representing the depleting SoC. This is reflected in a new set of directed arcs $\cA$ such that a node in a given layer is connected with a node in a layer below, i.e., with a lower SoC. In fact, nodes in the same layer are not connected because it would mean vehicles traveling without spending energy. As a consequence, the new set of arcs $\cA$ can only connect nodes in different layers so that the energy spent is reflected in an equivalent number of layers traversed down. %\msmargin{Note that this procedure leads to a large number of layers, which reflects in a large number of arcs and nodes of the multi-layer network. }{reframe it..sounds a bit negative}

\subsection{Travel Requests and Charging Stations}
On top of the multi-layer network, %$\mathcal{G} = (\mathcal{V}, \mathcal{A})$, 
we introduce a set of geo-nodes $\mathcal{V_\mathrm{G}}$. Every geo-node is connected via geo-arcs $\mathcal{A_\mathrm{G}}$ to every other node that represents the same geographical location, across all the battery SoC layers. Geo-arcs have zero weight because they connect nodes that represent the same geographical location. We define as $\mathcal{G} = (\mathcal{V}, \mathcal{A})$ the multi-layer network comprehensive of geo-nodes and arcs.
Then, we define $\mathcal{M} = \{1, ..., M\}$ as a set of travel requests. Each request $m \in \mathcal{M}$ is defined by a tuple ${r}_m = (o_m, d_m, \alpha_m) \in \mathcal{V_\mathrm{G}} \times \mathcal{V_\mathrm{G}} \times \mathbb{R}^+$ in which $\alpha_m$ is the number of users traveling from the origin $o_m$ to the destination $d_m$ per unit time. Note that travel requests are initialized on $\cV_{\mathrm{G}}$.

The user flow induced by each demand $m$ is defined as $x_{ij}^{m}$, with $ m \in \mathcal{M}$ and $(i, j) \in  \mathcal{A}$.
All user demand flows $x_{ij}^{m}$ are fulfilled by vehicles, and the rebalancing vehicle flows are defined as $x_{ij}^{\mathrm{r}}$, which we define as the flow on arc $(i,j) \in \mathcal{A}$ of vehicles without users on-board. Both the user demand flow and the rebalancing flow originate and terminate in a geo-node in the set $\mathcal{V_\mathrm{G}}$. 
We define a set of potential charging station locations as $\cC \subset \cV_\mathrm{G}$, with charging station $c \in \cC = \{1,2,..,\abs{\cC}\}$.  We define a vector with binary entry $\kappa_c$ indicating with $\kappa_c=1$ if there is a charging station at node $c$, and $\kappa_c=0$ otherwise. A charging station $c$ is modeled as a set of charging arcs $\mathcal{A}_\mathrm{S}^c$, which allows the vehicles inside a charging station $c$ to move from a layer to the subsequent layer above,  while remaining at the same geographic location. Vehicles can only be charged while rebalancing via charging arcs $(i,j) \in \mathcal{A}_\mathrm{S}^c, \forall c \in \cC$.
Hence, we extend the multi-layer network $\cG$ by including the set of charging arcs $(i,j) \in \mathcal{A}_\mathrm{S}^c$. % the set of arcs $\mathcal{A} = \mathcal{A_\mathrm{R}} \cup \mathcal{A_\mathrm{S}} \cup \mathcal{A_\mathrm{G}}$, and the set of vertices $ \mathcal{V} = \mathcal{V_\mathrm{R}} \cup \mathcal{V_\mathrm{G}}$.

\subsection{Pruning of the Multi-layer Network with Iso-energy Arcs}\label{sec:pruning}

While vehicles travel they also deplete SoC, meaning that they have to traverse down an equivalent amount of energy w.r.t.\ the distance driven. However, due to the discretization of the SoC, it is not guaranteed to have an exact match between distance traveled and energy used. In particular, given the wide range of arc lengths and their relative energy consumption, the SoC discretization %equal to the maximum common factor among all the . The procedure allows for the discretization factor to be set equal to the value of the weight of the arcs. Consequently, the number of layers becomes equal to the battery capacity divided by the discretization factor. 
should be extremely fine, i.e., the number of layers should be very large, to avoid such mismatches. Crucially, the complexity of a traffic flow model grows with the square of the number of nodes. In addition, when jointly optimizing for the siting of the charging infrastructure, the combinatorial nature of the formulation can potentially lead to either intractable problems or the necessity of large computational resources. Moreover, it is desirable to have a formulation that allows for the computation of the solution in a reasonable time period. This is the case when a large design space of the base vehicle has to be explored and tested. Thus, we devise an algorithmic procedure to transform the original road network into a new synthetic network with iso-energy arcs. In other words, the new synthetic network has the characteristic of having arcs with equal weight in terms of energy consumption, while minimizing the absolute error of the shortest path distance w.r.t.\ the original network. The full procedure can be found in~\ref{app:pruning}.
The contribution of this procedure is twofold: i) it decreases the number of nodes in every layer, while minimizing the errors caused by the pruning; ii) the constant weight allows for an exact match between distance traveled and energy consumed and consequently a reduction in the number of SoC layers needed. 
%A schematic representation of the first layer of the network is shown in Fig.~\ref{fig:FinalGraphs} for Manhattan, NYC. 

\subsection{Network Flow Model Formulation}
We set as objective the minimization of the number of vehicles, which is equivalent of minimizing the total travel time of the fleet~\citep{Pavone2015}:
\begin{equation}\label{eq: objective1}
    \min_{{x}^{m},{x}^\mathrm{r}}\sum_{{m} \in \mathcal{M}} \sum_{{(i,j)}\in\mathcal{A}} t_{ij}\cdot (x_{ij}^{m} + x_{ij}^\mathrm{r}),
\end{equation}
where $x_{ij}^{m}$ is the user demand flow, $x_{ij}^\mathrm{r}$ is the rebalancing flow, $t_{ij}$ is the time to traverse arc ${(i,j)}\in\mathcal{A}$. Note that this objective is equivalent to the minimization of the number of vehicles required to implement the obtained flows, as demonstrated by~\cite{PavoneSmithEtAl2012}.
The vehicle and user flow conservation are expressed as in a multi-commodity transportation problem
\begin{multline}\label{eq: Flow conservation}
        \sum_{(i,j)\in\mathcal{A}}x_{ij}^{{m}} + \mathds{1}_{j=o_{m}}\cdot \alpha_{{m}} = \sum_{(j,k)\in\mathcal{A}}x_{jk}^{{m}}+ \mathds{1}_{j=d_{m}}\cdot \alpha_{{m}}\\ \qquad \forall {m}\in\mathcal{M},
\end{multline}
where the user flow $x_{ij}^{m}$ is induced by demand $m$, $\mathds{1}_{x=y}$ is the indicator function, equal to $1$ if $x=y$ and zero otherwise, and $\alpha_m$ is the user request rate per unit time.
Rebalancing the vehicles in the E-AMoD system is critical to create a balanced system and to realign vehicle distribution with transportation requests. This is ensured by
\begin{equation} \label{eq: Flow Rebalance}
        \sum_{(i,j)\in\mathcal{A}} \left( x_{ij}^{\mathrm{r}} + \sum_{{m} \in \mathcal{M}} x_{ij}^{m}   \right) = 
         \sum_{(j,k)\in\mathcal{A}} \left( x_{jk}^{\mathrm{r}} + \sum_{{m} \in \mathcal{M}} x_{jk}^{m}   \right).
\end{equation}
The geo-nodes in $\mathcal{V}_\mathrm{G}$ act as origins and destinations for the vehicles. Consequently, each demand will require the geo-arcs $\mathcal{A}_\mathrm{G}$ to be used twice: first, from a geo-node to the road network, and then from the road network to a geo-node. The same is also applicable to rebalancing flows. This is ensured by constraining the user flow by
\begin{equation}\label{eq: FromGeo_vehicle}
    \begin{aligned}
        \sum_{{m \in \mathcal{M}}} \left(\sum_{(i,j) \in \mathcal{A}_\mathrm{G}} x_{ij}^{m}+\sum_{(k,l) \in \mathcal{A}_\mathrm{G}} x_{kl}^{m} \right) = 2 \cdot \sum_{{m}\in\mathcal{M}} \alpha_{{m}} \\ \qquad \forall {i,l} \in \mathcal{V}_\mathrm{G},
    \end{aligned}
\end{equation}
and the rebalancing flows to
\begin{equation}\label{eq: FromGeo_rebalance}
    \begin{aligned}
        \sum_{(i,j) \in \mathcal{A}_\mathrm{G}} x_{ij}^\mathrm{r}+\sum_{(k,l) \in \mathcal{A}_\mathrm{G}} x_{kl}^\mathrm{r} = 2 \cdot \sum_{{m}\in\mathcal{M}} \alpha_{{m}}  \qquad \forall {i,l} \in \mathcal{V}_\mathrm{G}.
    \end{aligned}
\end{equation}
We enforce SoC conservation when passing through a geo-node $\mathcal{V}_{\mathrm{G}}$ as
%Equation~\eqref{eq: vehicleRebalance_toGeo} ensures the same, but for demand flow going to the GeoNode, and rebalance flow from the GeoNode. 
\begin{equation}\label{eq: vehicleRebalance_Geo}
    \begin{aligned}
        \sum_{(i,j) \in \mathcal{A}_\mathrm{G}} x_{ij}^{m} - x_{ji}^\mathrm{r} = 0 \qquad \forall {m}\in\mathcal{M}, \forall {i,j} \in \mathcal{V}_\mathrm{G}.
    \end{aligned}
\end{equation}
We also impose the set of possible charging stations locations $C=\cV_{\mathrm{G}}$ and set the limit of the number of charging stations to $N\in\sN$ with
\begin{equation}\label{eq: sumCP}
    \begin{aligned}
        \sum_{c=1}^{\abs{\cC}} \kappa_{c} \leq N.
    \end{aligned}
\end{equation}
Each charging station has a limited capacity in terms of the number of vehicles it can charge simultaneously. The capacity constraint of each charging station is given by
\begin{equation}\label{eq: chargingArcs}
    \begin{aligned}
    \Delta E \cdot \sum_{(i,j) \in \cA_\mathrm{S}^c} x_{ij}^{\mathrm{r}} \leq  P_\mathrm{max} \quad \forall c \in \cC,
    \end{aligned}
\end{equation}
where $P_\mathrm{max}$ is the charging station power, and $\Delta E$ is the difference in SoC between the layers, which is a constant and which limits the rebalancing flow through the charging arcs ${(i,j)} \in \mathcal{A}_\mathrm{S}^c$. 
Since the vehicles must not charge with users on-board, the charging of the vehicles can be carried out only while rebalancing. This is ensured by
\begin{equation}\label{eq: Xv(charging) = 0}
    \begin{aligned}
        x_{ij}^{{m}} = 0 \qquad \forall {m}\in\mathcal{M}, \forall {(i,j)} \in \mathcal{A}_\mathrm{S}^c,
    \end{aligned}
\end{equation}
which guarantees that the vehicles do not charge when serving users. Therefore, the vehicles can only charge while rebalancing through charging arcs $(i,j) \in \mathcal{A}_\mathrm{S}^c$.
Then, we impose non-negative flow constraints,
\begin{align}
&x_{ij}^\mathrm{{r}} \geq 0 \qquad \forall {(i,j)} \in \mathcal{A}, \label{eq: Xv > 0} \\
&x_{ij}^{{m}} \geq 0 \qquad \forall {m}\in\mathcal{M}, \forall {(i,j)} \in \mathcal{A}.  \label{eq: Xu > 0}
\end{align}

To conclude, we define the E-AMoD optimization problem as follows:
\begin{prob}(E-AMoD Joint Optimization Problem)\label{prob:one}
	Given a set of transportation requests $\mathcal{M}$, the optimal user flows $x^m$, the rebalancing flows $x^r$, and the optimal siting of the charging infrastructure $c$, result from
	\begin{equation*}
		\begin{aligned}
			&\!\min_{{x}^{m},{x}^\mathrm{r},\kappa}\sum_{{m} \in \mathcal{M}} \sum_{{(i,j)}\in\mathcal{A}} & & t_{ij}\cdot (x_{ij}^{m} + x_{ij}^\mathrm{r}), \\
			& \textnormal{s.t. } & &\eqref{eq: Flow conservation}-\eqref{eq: Xu > 0} .
		\end{aligned}
	\end{equation*}
\end{prob}
Problem~\ref{prob:one} is a mixed-integer linear program (MILP) that can be solved with global optimality guarantees with off-the-shelf solvers. In addition, given a pre-defined set of charging stations, the problem can be cast as a linear program  that can be efficiently solved with interior-point algorithms in polynomial time.

\subsection{Ride-pooling Network Flow Model Formulation}
\begin{figure}[t]
	\centering
	\begin{subfigure}{\linewidth}
		\centering
		\includegraphics[width=.7\linewidth]{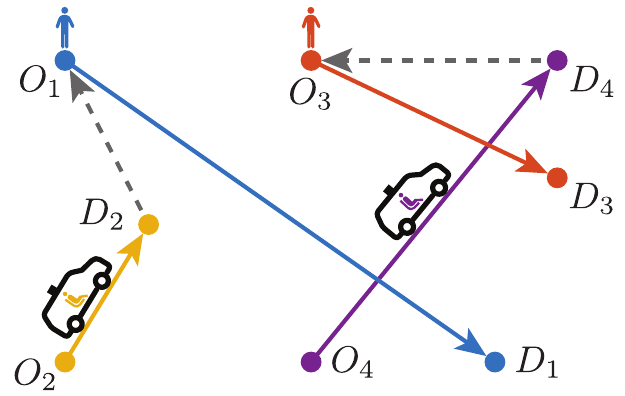}
		\caption{No ride-pooling.}
	\end{subfigure}\\
	\begin{subfigure}{\linewidth}
		\centering
		\includegraphics[width=.7\linewidth]{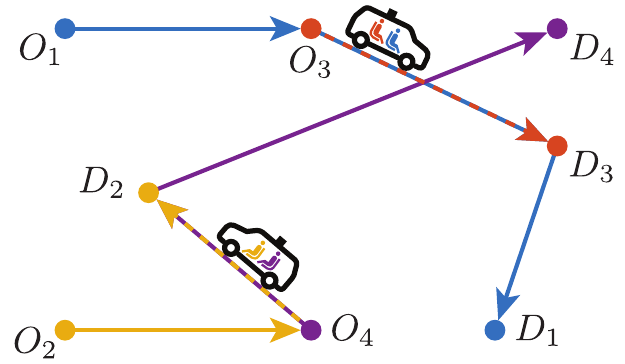}
		\caption{Ride-pooling.}
	\end{subfigure}
	\caption[]{Illustrative example of no ride-pooling and ride-pooling with two vehicles; dashed arrows indicate vehicles traveling without users on-board. Figure taken from~\cite{PaparellaPedrosoEtAl2024b}.}
	\label{fig:eg_ridepooling}
\end{figure}

In the previous section, the standard network flow model formulation was presented. One of the main assumptions of the model is that each vehicle can only transport one user at the time. However, the replacement of private vehicles with an AMoD fleet can lead to an increase in VHT due to the rebalancing term. Thus it is crucial to include ride-pooling, which is schematically depicted in Fig.~\ref{fig:eg_ridepooling}, allowing two or more users to travel on a vehicle at the same time. Given the combinatorial nature of the problem, it is important to find a method that is reasonably fast, in line with the method of this work. 
%First we notice that, following~\cite{Rossi2018}, it is possible to decouple the computation of the rebalancing flow from the user flow. As a \msmargin{consequence}{?}, 
Given a set $\cM$ of travel requests in a time-invariant network flow model, it is possible to compute an equivalent set of travel requests $\cM'$ that includes the possibility for users to ride-pool, which is optimal w.r.t.\ the VHT by the fleet with users on-board. In particular, the new set can be computed for as a function of the maximum waiting time $\bar{t}$, a maximum delay $\bar{\delta}$, and a maximum vehicle capacity $K$, which are the maximum waiting time a user is willing to wait another user, the maximum time caused by the detour w.r.t.\ no ride-pooling, and the maximum number of users that can be pooled together in the same vehicle at the same time, respectively. Given the new set $\cM'$, it is possible to solve Problem~\ref{prob:one} while taking into account ride-pooling. The idea has been proposed in~\cite{PaparellaPedrosoEtAl2024b}, which we refer to the  reader for a detailed analysis. In what follows, we briefly explain the methodology.
For the sake of simplicity, we restrict the number of users that can ride-pool together to $K=2$. Nonetheless, the reasoning can be applied to any $K$, as detailed in the aforementioned work. Given three travel requests $r_1 = (o_1,d_1,\alpha_1)$, $r_2 = (o_2,d_2,\alpha_2)$ and $r_3 = (o_3,d_3,\alpha_3)$, there are nine ways in which it is possible to ride-pool such travel requests, corresponding to the combinations between them (with repetitions and with permutations). In each ordered combination there is one way to pick up travel requests and one way to drop them off that minimizes the vehicle hour traveled by the fleet. For example, to ride-pool $r_1$ and $r_2$ (in that order), a vehicle should travel through the nodes $o_1,o_2$ in that order and then, starting from $o_2$, to $d_1,d_2$ in the order that minimizes the VHT. To force the vehicle to travel through those nodes, the ride-pooling travel request between $r_1,r_2$ can be represented by an equivalent set composed by three travel requests that would force a vehicle to first pick-up both the requests and then drop them off. An example would be:
\begin{equation*}
	r_{12}^1 = (o_1,o_2,\min(\alpha_1,\alpha_2)P(\alpha_1,\alpha_2,\bar{t}))
\end{equation*}
\begin{equation*} 
	r_{12}^2 = (o_2,d_1,\min(\alpha_1,\alpha_2)P(\alpha_1,\alpha_2,\bar{t}))
	\end{equation*} 
\begin{equation*}
	r_{12}^3 = (d_1,d_2,\min(\alpha_1,\alpha_2)P(\alpha_1,\alpha_2,\bar{t})), 
\end{equation*}
where $r_{12}^1,r_{12}^2,r_{12}^3$ are the equivalent travel requests that would force the vehicle to pass through $o_1,o_2,d_1,d_2$. The $\min(\cdot)$ imposes that the same number of users from both travel requests are taken, the function $P(\alpha_1,\alpha_2,\bar{t})$ is a probability function of having one single user from the travel request $r_1$ and one from $r_2$ within a time window $\bar{t}$ and it is given, for a generic number of $k$ travel requests, by 
\begin{equation}\label{eq:lem}
	P_{\bar{t}}\left(\alpha_{1}, \ldots, \alpha_{k}\right)=\sum_{i=1}^{k} \frac{\alpha_{i}}{\sum_{j=1}^{k} \alpha_{j}} \prod_{\substack{j=1\\j\neq i}}^{k}\left(1-e^{-\alpha_{j} \bar{t}}\right).
\end{equation}
Thus, for a given $r_1,r_2$ the equivalent set of ride-pooling travel requests composed by $r_{12}^1,r_{12}^2,r_{12}^3$ can be fully defined. Note that the probability given by ~\eqref{eq:lem} is always strictly lower than 1, meaning that two travel requests cannot be fully ride-pooled together, i.e., some users cannot be ride-pooled due to time constraints. The remaining part of such users that could not be pooled can, however, potentially be pooled with other travel requests. In the previous example, the remaining users from $r_1,r_2$ can be pooled with the users from $r_3$.
This second step is dependent on the previous step, as it depends on how many users remain unmatched from the original travel requests $r_1,r_2$ after the previous step. In addition, the order of matching $r_1,r_3$ and $r_2,r_3$ influences the final result. The priority has to be optimized accordingly to minimize the objective, which in this case is the VHT with users on-board. This procedure leads to a non-linear mixed integer problem, which can quickly become untractable, even for a small number of travel requests. 
To solve this problem in an efficient way, we leverage a Knapsack-like algorithm, which has been proved to be optimal w.r.t.\ the VHT by the fleet with users on-board~\citep[Theorem 1]{PaparellaPedrosoEtAl2024b}. The algorithm allows to prioritize the ride-pooling matches and allows to solve the problem in polynomial time. The equivalent set of travel requests obtained represents the expected number of ride-pooling matches that can be performed under maximum waiting time $\bar{t}$ and detour $\bar{\delta}$ constraints. Problem~\ref{prob:one} with the new set $\cM'$ allows to solve the minimum VHT problem.

\subsection{Discussion}
A few comments are in order. 
First, we model the system at steady-state. This assumption holds if the rate change of requests is significantly lower than the average travel time of individual trips, as observed in densely populated urban environments by \cite{Neuburger1971}. In addition, vehicle conservation implicitly enforces SoC conservation over the time-span under consideration.
Second, the iso-energy synthetic network procedure is only valid for environments where traveling to and from a node requires approximately the same amount of energy, e.g., flat urban environments. In the future, the authors intend to extended this model to capture more general (e.g., hilly) scenarios as well. 
Third, the results obtained by solving Problem \ref{prob:one} allow for fractional flows to occur. This is acceptable given the mesoscopic nature of the problem, where arc flows are in the order of hundreds of vehicles, see~\cite{LukeSalazarEtAl2021}. However, it is possible to retrieve near-optimal integer flows from fractional ones leveraging randomized sampling methods, as shown by~\cite{RossiIglesiasEtAl2017EV}. 
Fourth, we do take into account the exogenous traffic and its stochastic nature for a specific traffic scenario. In case of different conditions, the travel time and energy consumption can be updated accordingly, and a new synthetic network can be generated. Finally, we neglect the effect of endogenous traffic and the reaction of private vehicles. However, it is possible to readily account for exogenous traffic congestion by modifying the entries $t_{ij}$ accordingly.

\section{Results} \label{sec: Results}
We investigate two case studies for the metropolitan region of Manhattan, NYC. The original data set is extracted from OpenStreetMap, whilst to create the synthetic network we leverage the algorithm explained in Section~\ref{sec:pruning}. Following the results in~\ref{app:pruning}, we set the target weight to $\unit[100]{Wh}$ and the compression rate to $90\%$. 
The travel requests are publicly available from the website of the New York Taxi commission.
The data set used as a reference is the peak-hour from $8$ to $9$AM during March 1st, 2022, which has approximately $14'000$ travel requests. %Due to the high volume and the absence of strong peak hours of daily travel requests in NYC, see ~\cite{Meyers2018}, it is reasonable to assume that travel requests change slowly w.r.t\ the average travel time, allowing to not lose the assumptions of the linear time-invariant model. 
%Moreover, the results that will be shown in Section~\ref{sec: caseStudy1}, specifically the ratio between rebalancing and overall distance driven, are in line with~\cite{HogeveenSteinbuchEtAl2021}, enforcing that the linear time-invariant hypothesis over a day holds.
We investigate two case studies. The first one assesses the advantages of jointly optimizing the infrastructure siting w.r.t.\ a heuristic placement based on geo-nodes centrality. In particular, we analyze the impact of the charging \mbox{infrastructure} density and siting on the resulting rebalancing flow energy consumed by the whole fleet. Then, we investigate the benefit that ride-pooling has on sizing of the charging infrastructure. 
The second case study evaluates the performance of the E-AMoD system when different designs of electric vehicles int terms of battery capacity, passenger seats, and efficiency per unit distance driven are employed, see Table~\ref{table:Vehicle}.
\begin{comment}

\begin{table}[t]
	\centering
	\caption{User Requests (Manhattan)}
	\label{tab:demands}
	\begin{tabular}{l|l|l}
 Parameter		&  Value & Unit\\ % Eindhoven &
		\hline
	 %   Peak hour  &  8000 & \unit{requests/hour} \\ %1200 &
% 10 Days demands & 12 & \unit{million requests/hour}\\ % 400 &
		Daily demands &  120k &  \unit{requests/day}\\ %18000 &
			Selected area  &  54 & \unit{km^2} \\ %12.81 &
	\end{tabular}
\end{table}
\end{comment}
\begin{table}[t]
\begin{center}
\caption{Normalized Parameters of the Vehicles under Consideration.}
\label{table:Vehicle}
\begin{tabular}{ l|l|l|l }
% Parameter & \multicolumn{3}{c|} {Vehicle} & Unit\\
 
  & 2-seater & 4-seater & 4-seater$+$  \\ 
  \hline
  %Efficiency (WLTP) & 100.5 & 110.2 & 118.9  \\
  Energy consumption & 85\% & 93\% & 100\%  \\
  Battery capacity & 25\% & 60\% & 100\%  \\
  Mass & 69\% & 85\% & 100\% \\
  Number of Seats & 2 & 4 & 4
\end{tabular}
\end{center}
\end{table}
The three vehicles are a small 2-seater, which has the smallest battery capacity and the lowest energy consumption, a medium 4-seater, with a larger battery capacity and a higher energy consumption, and a 4-seater$+$, which has has the highest battery capacity, making it the heaviest and most energy-consuming one.
%\msmargin{The complete graph representation of the E-AMoD system, taking into account the SoC layers, consists of 430 SoC layers, 8170 nodes and 83350 arcs.}{aggiungi queste info per tutti i veicoli nella tabella}
%Problem~\ref{prob:one} was solved using the solver Gurobi 9.5, see \cite{GurobiOptimization2021}, on an Intel core i7-10850H, 32GB RAM in approximately $\unit[15]{min}$. This highlights that the creation of the synthetic network is a necessary step to obtain a tractable problem and compute a solution in a reasonable time.  

\subsection{Charging Infrastructure Design}\label{sec: caseStudy2}
\begin{figure}[t]
	\centering
	\includegraphics[width=\columnwidth]{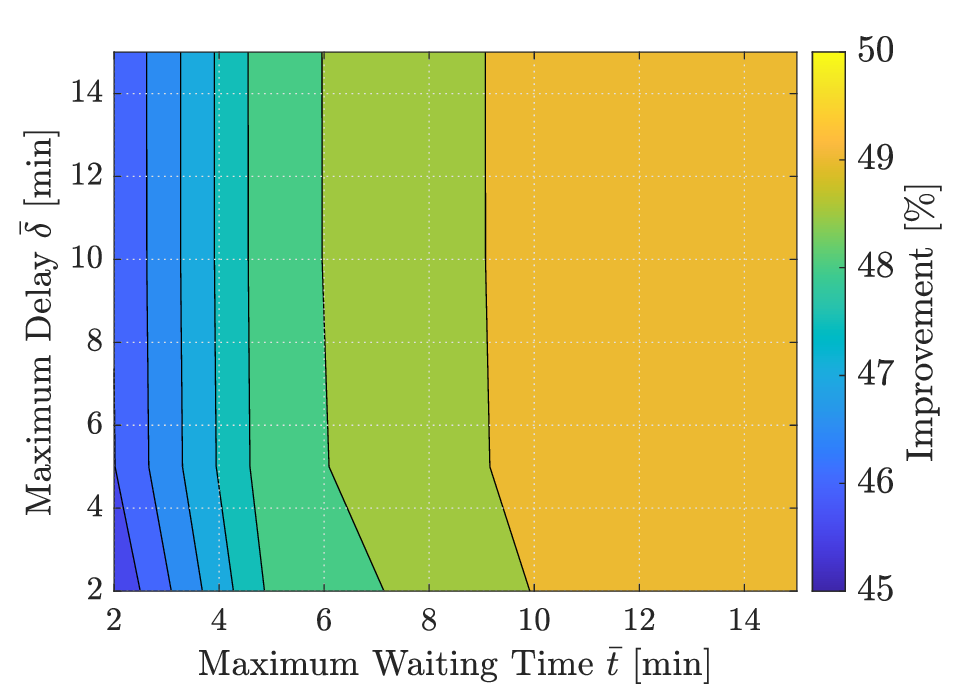}
	\caption{Improvement of overall daily energy usage of the ride-pooling scenario w.r.t.\ no pooling, for a varying waiting time and delay $\bar{t}$ and $\bar{\delta}$, respectively. Seat capacity of the vehicles $K=2$. Overall number of demands equal to $14$ thousand per hour. All charging stations have the same power available. }
	\label{fig:improv}
\end{figure}

In this case study we investigate the impact of the charging infrastructure density, siting, sizing and of ride-pooling on the energy consumption of a fleet of 4-seaters, but where we restrict the maximum number of users that can be ride-pooled together to $K=2$.
%\msmargin{It is important to remark that when dealing with the placing of the charging infrastructure, a long time period should be taken into account, in conjunction with travel requests uncertainty. This can be done by taking a larger time window and by including uncertainty in the travel requests, as explained by~\cite{BoydVandenberghe2004} in Ch. 4.}{?} This approach is similar to our method since the intrinsic nature of network flow models is equivalent to considering the expected value of the travel requests in a given time period.
We assess the advantages of jointly optimizing siting and routing by comparing the results of Problem~\ref{prob:one} w.r.t.\ the same framework, but where the siting of the charging infrastructure is selected a priori leveraging an heuristic approach. Namely, we use the \textit{betweenness} centrality,~\cite{Bullo2018}, i.e., we place the charging stations in the nodes with the highest probability of appearing on the shortest path between any two random nodes, as shown in Fig.~\ref{fig:FinalGraphs}.
As an alternative, other methods like k-mean clustering could be implemented.
\begin{figure}[t]
	\centering
	\includegraphics[trim={13 2 30 10},clip,width=\columnwidth]{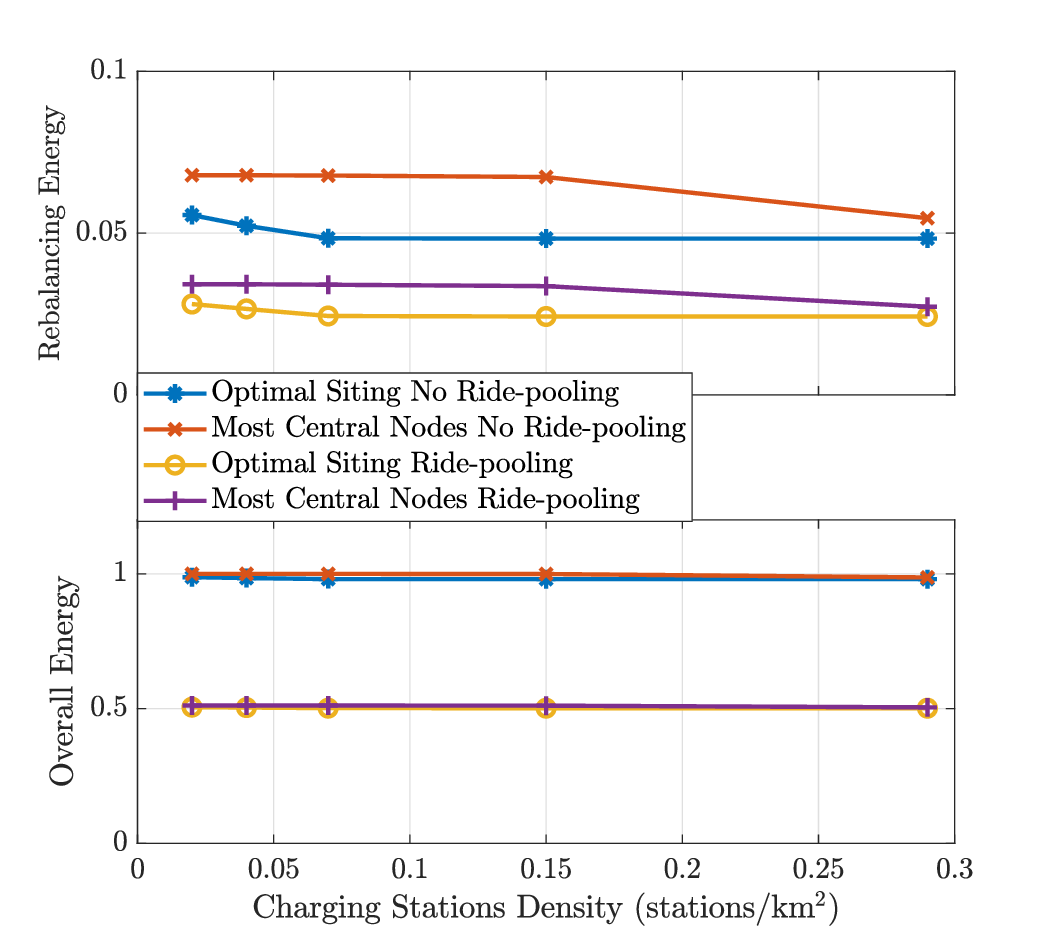}
	\caption{The rebalancing and overall energy consumption for the whole fleet leveraging an optimal approach and an heuristic method, betweenness centrality, to do the siting of the charging infrastructure for Manhattan, NYC. Results are both with and without ride-pooling scenario, with a maximum of two users per vehicle at the same time, $K=2$. The results are normalized w.r.t.\ the overall energy usage of the no ride-pooling scenario with heuristic placement. }
	\label{fig:Rebal}
\end{figure}
Fig.~\ref{fig:Rebal} shows the normalized overall and rebalancing energy usage in four scenarios for a varying density of charging stations per unit area. The four scenarios are with and without ride-pooling, and with optimal and heuristic placement. For the ride-pooling scenarios we set the maximum waiting time and the maximum delay of users to $\bar{t},\bar{\delta}=\unit[10]{min}$, and the maximum number of users that can ride-pool together to $K=2$. We recall that the waiting time is the time that a user is willing to wait for another user to be pooled together, and delay is the additional detour time w.r.t\ no pooling.
We notice that when going for the optimal siting of the charging infrastructure, it is possible to decrease the overall energy consumption of the fleet by approximately $1\%$ compared to heuristic placement in both the cases of ride-pooling and no ride-pooling. Such improvement is significantly lower compared to the one given by allowing for ride-pooling. In fact, given that users can share vehicles at the same time, a lower number of vehicles is required and the overall total energy is significantly reduced. When comparing rebalancing energy consumption per vehicle, the results are very similar between ride-pooling and no ride-pooling. In addition, when performing ride-pooling, the optimal placement remains the same, and  the rebalancing flow patterns do not vary significantly. As a result, when designing the siting of the charging infrastructure in a ride-pooling framework, it is possible to leverage state-of-the-art methodologies for mobility systems that do not account for ride-pooling. The results show that above a density of $\unit[0.1]{stations/km^2}$, if the siting is optimal, we reach a plateau. Up to this point, assuming each station is large enough to always have free charging spots, it is not convenient to build additional stations. On the contrary, by using the betweenness centrality heuristic method, the number of charging stations needs to increase to obtain a similar performance. %We highlight that these results are partially influenced by the granularity of the network, which is influenced by the available computational power. 
%Moreover, that siting is of secondary importance compared to sizing, which is the crucial factor that allows for the correct operation of the fleet and which might impose a higher number of charging stations if they have a maximum limit on the power output. 
In contrast with siting, sizing is strongly dependent on the decision of allowing for ride-pooling or not. Fig.~\ref{fig:improv} shows the improvement of overall energy usage of ride-pooling w.r.t.\ no ride-pooling. Interestingly, increasing the delay does not significantly improve the energy consumption, as the vehicles perform more detours to pick-up and drop-off users. On the contrary, waiting time strongly impacts energy consumption because users and vehicles waiting implies no energy consumption. The results show that ride-pooling can lead to a significant decrease in energy usage, in the order of $45\%$, even for small maximum waiting time and delay experienced by users. This result is mainly due to the very large number of travel requests that occur in Manhattan, which allows for an efficient matching of users. However, different regions with lower number of travel requests will experience a lower improvement. We refer to~\cite{PaparellaPedrosoEtAl2024b} for a detailed analysis on the correlation between improvement in VHT, which can be directly correlated to energy consumption and number of travel requests. 

\subsection{Impact of the Vehicle Design on the Optimal Operations}\label{sec: caseStudy1}
\begin{figure}[t]
	\centering
	\includegraphics[trim={10 2 10 10},clip,width=\columnwidth]{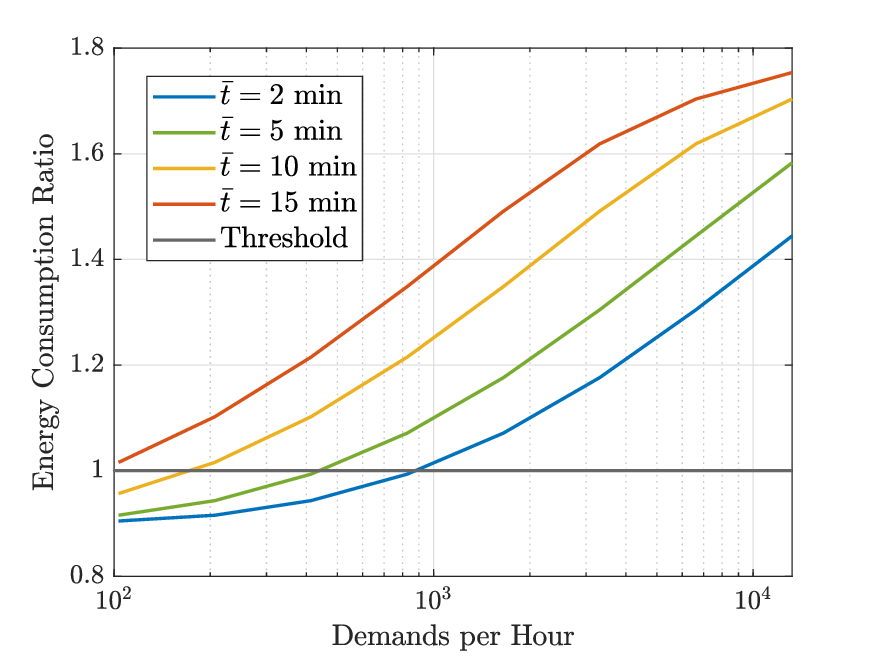}
	\caption{Energy consumption ratio between a fleet of 2-seaters and 4-seaters, with a maximum delay set to $\bar{\delta} =  \unit[10]{min}$. Above 1000 demands per hour, using heavier 4-seater is more energy efficient even for low waiting times $\bar{t}$ of \unit[2]{min} only.}
	\label{fig:ratio}
\end{figure}

This section shows a second case study where we assess the impact of vehicles' characteristics on the performance of the E-AMoD system. Specifically, we select three different commercial vehicles of which the normalized parameters are shown in Table \ref{table:Vehicle}. The parameters of these vehicles are normalized w.r.t.\ the LightYear One~\citep{Lightyear}. Following the results of Section~\ref{sec: caseStudy2}, we solve Problem~\ref{prob:one} with an optimally sited charging infrastructure of density equal to $\unit[0.1]{stations/km^2}$.
\begin{figure}[t]
    \centering
    \includegraphics[width=\columnwidth]{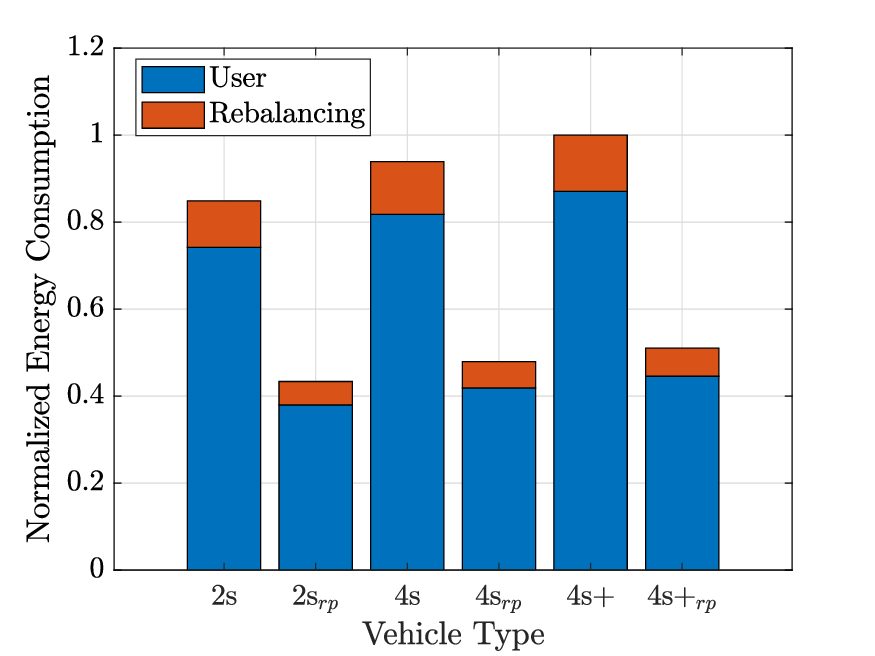}
    \caption{Energy used by vehicles with users on-board (blue) and rebalancing vehicles without users on-board (red) for 2-seater (2s), 4-seater (4s) and 4-seater$+$ (4s$+$) in Manhattan, with ride-pooling (\textit{rp}) and without it. Maximum waiting time and delay are set to $\bar{t},\bar{\delta} =  \unit[10]{min}$, respectively. The maximum number of users that can be ride-pooled together at the same time is $K=2$ for every vehicle. Number of travel demands equals to 14 thousand per hour.}
    \label{fig:CS1}
\end{figure}
We perform a simulation for every type of vehicle, both with and without ride-pooling with a maximum delay and waiting time of $\bar{\delta},\bar{t}=\unit[10]{min}$ and a maximum of two users per vehicle at the same time, $K=2$. Then, we analyze the differences in terms of the user flow and rebalancing flow energy consumption, as illustrated in Fig.~\ref{fig:CS1}.
We highlight that the rebalancing flow distance in the three cases with different vehicles is the same and that the difference in energy consumption is mainly due to the difference in consumption per unit distance driven. This result is in line with the results obtained by~\cite{HogeveenSteinbuchEtAl2021,PaparellaHofmanEtAl2024}, which show that in mobility-on-demand frameworks, vehicles with battery capacity above a certain threshold have a comparable rebalancing flow distance traveled (including detour to charging stations). 
First, our results indicate that when choosing vehicles for mobility-on-demand, it is important to give priority to efficiency in terms of consumption per unit distance driven compared to battery capacity. In fact, the majority of commercial vehicles have a battery capacity which is above a threshold, which results in approximately constant rebalancing distance.  
Second, we highlight that allowing for ride-pooling leads to a significant decrease in overall distance driven and energy consumed. This is especially the case in a densely populated urban environment where it is possible to efficiently and conveniently match together the majority of the travel requests. In what follows, we will compare the performance of the 2-seater and the 4-seater for a varying number of travel requests, and allowing users to be ride-pooled together up to the number of seats in each vehicle, i.e., $K=2,4$, respectively. We disregard the 4-seater$+$ as we have shown that the additional battery capacity does not lead to any improvement. Fig.~\ref{fig:ratio} shows the ratio of overall energy consumption between a fleet of 2-seaters and 4-seaters where it is possible to ride-pool users up to the number of seats in the vehicles. The results indicate that it is not always convenient to choose the smallest vehicles, but that the design choice is strongly influenced by both the number of travel requests and the maximum waiting time $\bar{t}$ that users are willing to withstand. In fact, for a low maximum waiting time $\bar{t}$ and a low number of travel requests, pooling more than two users is very unlikely and inefficient. As a consequence, the higher efficiency of the 2-seater  overcomes the drawbacks of having less user seats. Conversely, for a higher waiting time and number of travel requests, the higher energy consumption of a larger vehicle is counterbalanced by the possibility of efficiently pooling together more than two users.
To conclude, these results show that the best vehicle choice is strongly influenced by some parameters, like waiting time and number of requests, and that, in ride-pooling systems energy consumption can be outweighed by the number of user seats in the vehicle. In a densely populated area like Manhattan, where there is a large number of travel requests per unit time, it is convenient to have a larger vehicle with lower energy efficiency per unit distance but more seats, i.e., a 4-seater shall be preferred over a 2-seater. %In not populated areas, where it is unlikely to efficiently pool many users together, conversely, it shall be preferred vehicle A over vehicle B, as energy efficiency 
\section{Conclusion}\label{sec: Conclusions}
This paper proposed a modeling and optimization framework to jointly optimize operations and charging infrastructure design for Electric Autonomous Mobility-on-Demand (E-AMoD) systems with ride-pooling in a computationally tractable, scalable and globally optimal fashion.
We devised a time-invariant network flow model that accounts for the State of Charge (SoC) of the vehicles, without the actual fleet size affecting its computational complexity.
Leveraging this model, we formulated the ride-pooling E-AMoD joint operation and infrastructure problem as a mixed-integer linear program that can be solved with global optimality guarantees using off-the-shelf optimization algorithms. 
%\msmargin{To improve tractability, we devised a smart pruning algorithm that allowed to generate a synthetic road network with equal weights while minimizing the shortest path error w.r.t.\ the original road network.  }{not the main reason. can delete this}
%In particular, when showcasing our framework using real-world taxi data from Manhattan, NYC, the LP and MILP were solved in 5 and \unit[15]{min}, respectively.
Our case study showed that jointly optimizing the charging infrastructure placement optimally can improve the performance achievable by the E-AMoD system by approximately $1\%$.
%It also revealed that the benefit of increasing the number of charging stations vanishes rapidly in case of optimal siting of the infrastructure, but this might not occur if the placement is heuristic. 
In addition, we showed ride-pooling can significantly  %influence neither the rebalancing flows of the vehicles nor the optimal placement of the charging infrastructure, therefore it is possible to leverage state-of-the-art algorithms that do not take ride-pooling into consideration. Moreover, ride-pooling strongly influences the travel time of the fleet, and as a consequence, the sizing of the infrastructure.
decrease energy consumption, with an improvement of up to 45\% for densely populated urban environments.  
Lastly, we showed that the optimal choice of the vehicles is strongly influenced by the number of users in the systems and waiting time. For the case study of Manhattan, using 4-seaters appeared to be more beneficial albeit their higher energy consumption, as it allows to efficiently pool together more than two users.

For future work, we would like to take into account the total costs of ownership of the fleet, and the costs for the charging infrastructure and its capacity; study intermodal settings~\citep{SalazarLanzettiEtAl2019,Wollenstein-BetechSalazarEtAl2021}; and include incentives for ride-pooling~\citep{FielbaumKucharskiEtAl2022,PedrosoHeemelsEtAl2023}.
%Finally, it would be insightful to account for the interactions with the power grid and study heterogeneous electric fleet compositions.
%\msmargin{Note that this analysis is restricted to the energy usage of the fleet.	For the sake of completeness, the total cost of ownership shall be taken into account which can be composed by many factors like parking cost, purchasing cost, maintenance, vehicle life time and end of life retained value per vehicle.We leave this interesting topic to future research.}{move to a shorter sentence in the outlook?}
\section{Acknowledgments}\label{Sec:akn}
We thank Dr. I. New for proofreading the paper. This publication is part of the project NEON with number 17628 of the research program Crossover.
% References
\bibliographystyle{elsarticle-harv} 
\bibliography{main.bib,SML_papers.bib,mobility.bib}
\appendix
\section{Iso-energy Pruning of the Road Network}\label{app:pruning}
%\vspace{-5cm}
\begin{figure}[t]
	\centering
	\begin{tikzpicture}[->, >=stealth', every node/.style={scale=0.75}, node distance=2cm, semithick,
	state/.style ={draw, circle},
	ell/.style = {draw, ellipse, minimum height=1cm, minimum width=2cm},
	rect1/.style = {draw, rectangle, minimum height=0.75cm, minimum width=3cm},
	rect2/.style = {draw, rectangle, minimum height=1.15cm, minimum width=3cm, align=center},
	dia/.style = {draw, diamond, aspect=1.75}]
	
	\node[ell] (n1) at (0,0.2) {Start};
	\node[rect2] (n2) at (0, -0.95) {Compose list \\ of candidates};
	\node[rect1] (n3) at (0, -2) {Select candidate};
	\node[rect2] (n4) at (0, -2.9) {Shrink \& \\ Homogenization};
	%\node[rect1] (n4) at (0, -2.9) {Homogenization};
	\node[rect1] (n5) at (0, -3.8) {Evaluate error};
	\node[dia, align=center] (n6) at (3, -3.8) {Candidates \\ evaluated?};
	
	\node[rect2] (n7) at (0, -4.85) {Shrink best \\ candidate};
	\node[dia, align=center] (n8) at (0, -6.4) {Target size \\ reached?};
	
	\node[rect1] (n9) at (3, -6.4) {Homogenization};
	\node[ell] (n10) at (3, -7.4) {End};
	
	\path (n1) edge node {} (n2)
	(n2) edge node {} (n3)
	(n3) edge node {} (n4)
	(n4) edge node {} (n5)
	
	(n5) edge node {} (n6)
	
	(n7) edge node {} (n8)
	(n8) edge node [pos=0.35,above] {Y} (n9)
	(n9) edge node {} (n10);
	
	\draw [->] (n6.north) -- node[pos=0.75,left] {N} +(0,0.25) |- (n3);
	\draw [->] (n6.south) -- node[pos=0.75,left] {Y} +(0,-0.25) |- (n7);
	
	\draw [->] (n8.west) -- node[pos=0.5,above] {N} +(-0.5,0) |- (n2);
	
\end{tikzpicture}                
	\caption{Flow chart of the algorithm used to create the synthetic network in Fig.~\ref{fig:FinalGraphs}, where "Shrink" refers to the algorithm devised in~\cite{SadriSalimEtAl2017}.}
	\label{fig:flowc}
\end{figure}
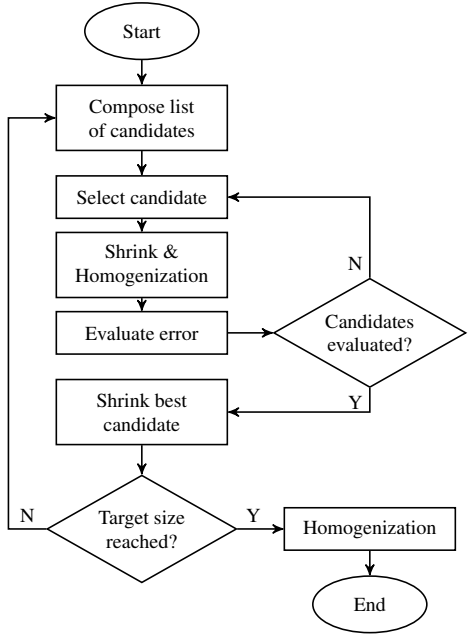 
\begin{figure}[t]
	\begin{subfigure}{0.49\linewidth}
		\centering
		\includegraphics[trim={0 0 00 20},clip,width=0.8\columnwidth]{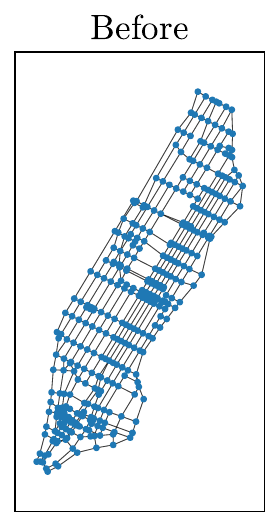}
		\caption{Original road network of Manhattan, NYC.}
		\label{fig:manhattan_before}
	\end{subfigure}
	\begin{subfigure}{0.49\linewidth}
		\centering
		\includegraphics[trim={0 0 00 20},clip,width=0.8\columnwidth]{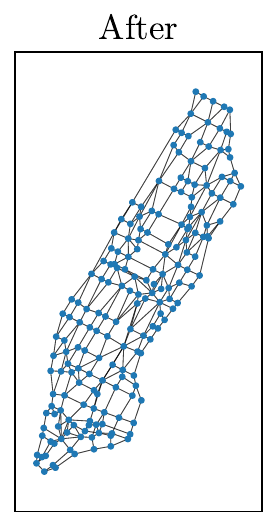}
		\caption{Synthetic road network of Manhattan, NYC for $w_c=\unit[50]{Wh}$.}
		\label{fig:manhattan_after}
	\end{subfigure}
	\caption{Original road network and synthetic one of Manhattan, NYC. In the synthetic network, each arc might be split into several arcs of weight equal to the target one. Such sub-division is not represented in the figure.}
	\label{fig:manhattan_graph}
\end{figure}
\begin{figure}[t]
\centering
\includegraphics[trim={0 0 00 0},clip,width=\columnwidth]{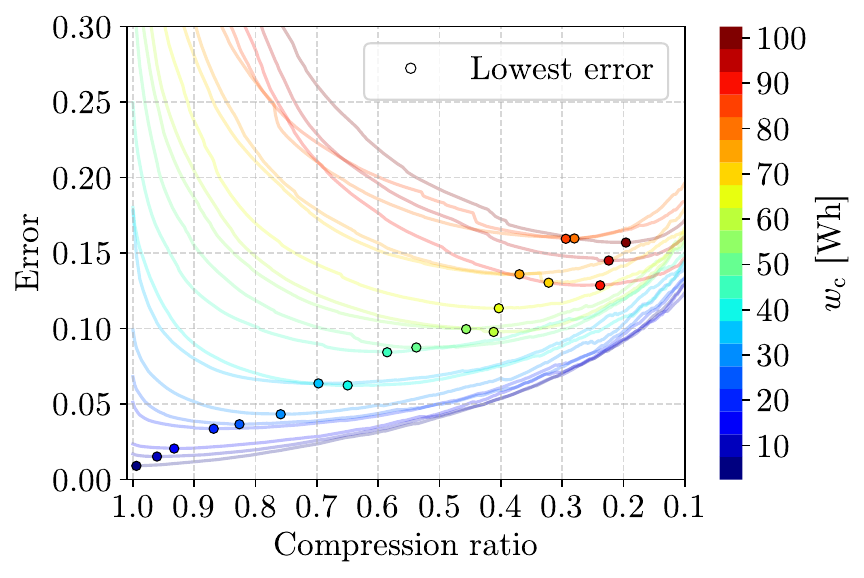}
\caption{Average absolute error of the shortest path between every node in the network as a function of the target weight and compression ratio. Results obtained for the network of Manhattan, NYC.}
\label{fig:scan}
\end{figure}
In this section we explain the algorithm used to create, starting from the original road network, a new synthetic road network with iso-energy arc weights in terms of energy consumption. We leverage the Shrink method devised by~\cite{SadriSalimEtAl2017}. Compared to the original formulation, we modify the algorithm by expanding the node selection procedure to every node and merging the two nodes such that the absolute difference between all the adjacent arcs w.r.t.\ the two nodes and the target weight (or a multiple integer of the target weight) is minimized. Fig.~\ref{fig:flowc} shows the flowchart of the algorithm based on Shrink.
We refer the reader to the paper written by~\cite{SadriSalimEtAl2017} for a detailed analysis of the algorithm.
Our results of Manhattan, NYC, show that when applying the algorithm described above, there is a correlation between the desired compression rate, the optimal target weight and the average absolute error of the shortest path for every couple of nodes. In particular, Fig.~\ref{fig:scan} shows the relation between the quantities mentioned previously, and highlights that for a given compression ratio, there is a target weight which minimizes the expected error. Fig.~\ref{fig:manhattan_graph} shows an example of the difference between an original network and a synthetic one generated with the method presented. Note that in the figure, each arc might be split into multiple arcs in series so that the total weight is an integer multiple of the target weight. Such arcs in series are not depicted in the figure.

%\section{Computation of Equivalent Ride-pooling Set}\label{app:ridep}

\end{document}